\documentclass[aps,pre,twocolumn,amsmath,amssymb,superscriptaddress,reprint,longbibliography]{revtex4-1}
\usepackage{amsfonts,amsmath,amssymb,bm}
\usepackage{graphicx,graphics,float}
\usepackage{bm}
\usepackage{amssymb}
\usepackage{colordvi}
\usepackage{graphicx}
\usepackage{color}
\usepackage[colorlinks=true,linkcolor=blue,citecolor=blue,urlcolor=blue]{hyperref}%
\usepackage{hyperref}
\usepackage{comment}
\usepackage{harpoon}
\usepackage{braket}

\begin{document}

\title{Geometric thermodynamic uncertainty relation in periodically driven thermoelectric heat engine}

\author{Jincheng Lu}\email{These authors contributed equally to this work.}
\address{Center for Phononics and Thermal Energy Science, China-EU Joint Lab on Nanophononics, Shanghai Key Laboratory of Special Artificial Microstructure Materials and Technology, School of Physics Science and Engineering, Tongji University, Shanghai 200092, China}

\author{Zi Wang}\email{These authors contributed equally to this work.}
\address{Center for Phononics and Thermal Energy Science, China-EU Joint Lab on Nanophononics, Shanghai Key Laboratory of Special Artificial Microstructure Materials and Technology, School of Physics Science and Engineering, Tongji University, Shanghai 200092, China}

\author{Jiebin Peng}
\address{Center for Phononics and Thermal Energy Science, China-EU Joint Lab on Nanophononics, Shanghai Key Laboratory of Special Artificial Microstructure Materials and Technology, School of Physics Science and Engineering, Tongji University, Shanghai 200092, China}

\author{Chen Wang}\email{wangchenyifang@gmail.com}
\address{Department of Physics, Zhejiang Normal University, Jinhua, Zhejiang 321004, China}

\author{Jian-Hua Jiang}\email{jianhuajiang@suda.edu.cn}
\affiliation{Institute of Theoretical and Applied Physics, School of Physical Science and Technology \&
Collaborative Innovation Center of Suzhou Nano Science and Technology, Soochow University, Suzhou 215006, China.}

\author{Jie Ren}\email{xonics@tongji.edu.cn}
\address{Center for Phononics and Thermal Energy Science, China-EU Joint Lab on Nanophononics, Shanghai Key Laboratory of Special Artificial Microstructure Materials and Technology, School of Physics Science and Engineering, Tongji University, Shanghai 200092, China}

\date{\today}
\begin{abstract}

Thermodynamic uncertainty relation,  quantifying a trade-off among average current, the associated fluctuation (precision), and entropy production (cost), has been formulated in nonequilibrium steady state and various stochastic systems. Herein, we study the thermodynamic uncertainty relation in   generic thermoelectric heat engines under a periodic control protocol, by uncovering the underlying Berry-phase-like contribution.
We show that our thermodynamic uncertainty relation  breaks the seminal steady-state results, originating from the non-vanishing geometric effect. Furthermore, by deriving the consequent trade-off relation binding efficiency, power, and constancy, we prove that the periodically driven thermoelectric heat engines can generally outperform the steady-state analogies. The general bounds are illustrated by an analytically solvable two-terminal single quantum dot heat engine under the periodic modulation. Our work provides a geometric framework in bounding and optimizing a wide range of periodically driven thermoelectric thermal machines.

\end{abstract}

\maketitle

\section{Introduction}

Periodically driven quantum machines reach  limited-cycle states after a long-time evolution since the coupling to the environment prevents infinite heating up~\cite{JungPR,fluctuationRMP,CampisiRMP,SeifertPR,PekolaPRM21}. These limited-cycle states form the basis of various functional thermal machines, which exhibit non-negligible fluctuations~\cite{JiangCRP,BenentiPR,Review21}. Investigating their trade-off relations provides insights into the optimal design principles for such periodically driven systems.

Recently, a thermodynamic uncertainty relation (TUR) has been formulated based on classical Markovian steady states,
 which demonstrates the trade-off relation between relative current fluctuation and dissipation~\cite{TUR15,HasegawaPRL,TURPRE19,TimpanaroPRL19,PietzonkaPRL,horowitz20,Falasco20,Dechant,LiuPRL20,SegalPRBTUR20,HasegawaPRL20,BijayPRR20,HartichPRL,HartichPRL,HasegawaPRL21,SegalPRE21}. Specifically, the average current $\langle I \rangle$, its variance $\langle\langle I^2 \rangle\rangle \equiv \braket{(I-\braket{I})^2}$, and the entropy production rate $\langle \sigma \rangle$ is universally bounded as
\begin{equation}
\frac{\langle\langle I^2 \rangle\rangle}{\langle I \rangle^2}\langle \sigma \rangle\ge2.
\label{eq:TUR}
\end{equation}
It is known that the TUR was initially proposed in the long-time limit~\cite{TUR15,GingrichPRL} and later generalized to the finite-time dynamics~\cite{pietzonka2017finite}. Consequently, the analysis methods and corresponding physical implications are further refined~\cite{MacieszczakPRL18,Proesmans19,BijayPRL21}.
Although TUR has been widely applied in tremendous amount of systems,  it is not always valid.
TUR violation corresponds to the situation that the left-hand side of Eq.~\eqref{eq:TUR} is smaller than $2$.
Involving quantum coherence~\cite{BijayPRB18}, temporal driving~\cite{Proesmans17, KoyukPRL19, SeifertPRL20}, and magnetic field breaking time-reversal symmetry~\cite{MacieszczakPRL18, saryal21} will violate the original TUR.

For periodically driven systems, inferring the entropy production or at least an upper bound, is generally more complex~\cite{BrandnerPRL17,KoyukPRL19,SeifertPRL20,VanVu20,BrandnerCoherent,TURPRE21,MillerPRL21,BrandnerPRX21,Menczel21}. An early counterexample  showed that a naive extension of the TUR from steady-state systems to periodically driven counterparts is inaccessible, since driving itself provides an spontaneous time scale enhancing current precision without significantly increasing the entropy production~\cite{BaratoPRX}. Subsequent attempts to find the analog for periodically driven systems yields Proesman and van den Broeck's bound, which is valid for time-symmetric drivings~\cite{Proesmans17}. Also, a series of general TUR incorporating driving speed's effect are proposed both for discrete and continuous state spaces~\cite{koyuk2018generalization, KoyukPRL19, SeifertPRL20}.

From the geometric view, the  underlying the state space has an intrinsic effect on periodically driven transports and time-dependent energy conversion processes.
Specifically, the geometric concepts are used in finite time thermodynamics~\cite{BrandnerPRX15,BrandnerPRL20}, in which the thermodynamic length~\cite{crooks2007measuring} bounds the engine power and efficiency. The Berry-phase-like effect provides an additional geometric contribution~\cite{brouwer1998scattering,NemenmanPRL, Sinitsyn07, RenPRL10,ChenPRB13, MillerPRL, wangpump, wang2021} to pump electric and heat currents  against the thermodynamic bias. However, the intrinsic effects of geometric phase on TUR and the performance of thermal machines are largely overlooked in the previous studies. To address the geometric effect in time-dependent system, in this work, we study the thermodynamic uncertainty relation, heat-work conversion, trade-off between energy efficiency, electric work, and work fluctuations in the periodically driven thermoelectric heat engine.
In particular, we find that non-vanishing geometric phase can simultaneously enhance the constancy of the engine, while not significantly introducing further entropy production.

We highlight the difference between our work and previous studies here. Firstly, our results are not restricted to time-symmetric driving protocols, in contrast to Ref.~\cite{Proesmans17}, which breaks down in general asymmetric protocols. Secondly, we unveil the role of geometry in driven systems, which stays unclear in previously derived TUR relations~\cite{koyuk2018generalization, KoyukPRL19,SeifertPRL20}. The geometric phase provides a versatile principle for implementing device design and optimization. Meanwhile, our framework can be readily generalized to more complex thermal machines possibly undertaking multi-tasks~\cite{manzano2020hybrid}.

The paper is organized as follows. In Sec. \ref{sec-bounds}, we organize
the TUR relationship while using the intrinsic geometric origin and analyze the bounds on electric work and energy efficiency in periodically driven thermoelectric heat engine. We study examples and verify the validity of the TUR using a two-terminal single quantum dot system in Sec. \ref{verify}. We conclude in Sec. \ref{conclusion}. Throughout, we set the Boltzmann constant to $1$.

\section{The universal bounds in periodically driven thermoelectric heat engine}~\label{sec-bounds}
\subsection{The bounds on fluctuations and entropy production}

For periodically driven systems with a period ${\mathcal T}\equiv 2\pi/\Omega$~\cite{KoyukPRL19}, Koyuk and Seifert derived a family of inequalities that relate entropy production with experimentally accessible data for the mean, its dependence on driving frequency, and the variance of a large class of observables,
\begin{equation}
\frac{\langle\langle I^2(\Omega) \rangle\rangle}{\langle I(\Omega) \rangle^2} \langle \sigma(\Omega) \rangle\ge 2\left[1 - \Omega\frac{d\langle I(\Omega)\rangle}{d\Omega} \frac{1}{\langle I(\Omega)\rangle} \right]^2.
\label{eq:TUR-I}
\end{equation}
The left-hand side involves the same combination of variables as the ordinary TUR does, where the dependence on $\Omega$ is explicit. The right-hand side additionally contains the derivative of the current with respect to the driving frequency, i.e., the response of the current to a slight change of the period of driving.

The current contribution composed of two parts: the dynamic current and the geometric one, i.e., $\braket{I}=\braket{I}|_{\rm dyn}+\braket{I}|_{\rm geo}$. The dynamic part is simply an average over instantaneous steady states, whereas the geometric part originates directly from the time dependence of the cyclic state. Near the adiabatic regime, $\braket{I}|_{\rm dyn}$ is independent of $\Omega$, and the geometric current $\braket{I}|_{\rm geo}$ is proportional to the driven frequency $\Omega$~\cite{YugePRB,WangPRA17}. The reason for this scaling will be explained in later sections.
Consequently, the right-hand side of Eq.~(\ref{eq:TUR-I}) can be simplified as
\begin{equation}
\begin{aligned}
1 - \Omega\frac{d\langle I(\Omega)\rangle}{d\Omega} \frac{1}{\langle I(\Omega)\rangle} = \frac{1}{1 + \langle I\rangle|_{\rm geo}/\langle I\rangle|_{\rm dyn}}.
\end{aligned}
\end{equation}
Thus, we arrive at the bound
\begin{equation}
\frac{\langle\langle I^2 \rangle\rangle}{\langle I \rangle^2} {\langle \sigma \rangle}\ge 2\left[\frac{1}{1 + \langle I\rangle|_{\rm geo}/\langle I\rangle|_{\rm dyn}} \right]^2\equiv \epsilon_{\rm bound}.
\label{eq:TURp}
\end{equation}
This is our first main result. In the adiabatic limit $\Omega \to 0$, we generally have $I|_{\rm geo}/I|_{\rm dyn} \to 0$. Therein, the geometric contribution is negligible, the dynamic part becomes dominant. Accordingly, the thermodynamic bound reproduces the ordinary TUR~\cite{TUR15}.
However, one consequence of this relation is that it provides a generic condition for (almost) dissipation-less precision. If the current is nearly proportional to the frequency of driving, where $I|_{\rm geo} \gg I|_{\rm dyn}$, the right-hand side vanishes. Temporally driven systems without the static bias lie in the possible implementations where this optimal limit can hold. This phenomenon is nonexistent at the steady states.

\subsection{The bounds on electric work and energy efficiency in periodically driven thermoelectric heat engine}


We consider a system isothermally coupled to several reservoirs with which it can exchange particle and energy. 
The total entropy production $\braket{\sigma}$ is specified by the stochastic thermodynamics~\cite{SeifertPR}
\begin{equation}
\label{eq:entropy production}
T\langle \sigma \rangle =-\langle W_{\rm out} \rangle+\langle W_d \rangle + \langle W_I \rangle.
\end{equation}
In the right-hand side, the first term denotes the output work $\langle W_{\rm out} \rangle$ (useful work), the second term $\langle W_d \rangle$ represents dissipation (dissipated work), and the last term is the input energy $\langle W_I \rangle$ (done by the temporal driving) accumulated over one period. $T$ is the temperature of the reservoirs. We restrict here to cyclic states, where the average entropy production of the middle system is zero in a full cyclic period. The positive energy is defined by flowing from the reservoirs into the system.

In this subsection, we consider the heat engine regime ($\braket{W_{\rm{out}}} >0$). Concentrating on the work fluctuations using Eq.~\eqref{eq:TUR-I}, we obtain
\begin{equation}
\frac{\langle\langle W_{\rm out}^2 \rangle\rangle}{\langle W_{\rm out} \rangle^2} {\langle \sigma \rangle}\ge 2\left[1 - \Omega\frac{d\langle W_{\rm out}(\Omega)\rangle/d\Omega}{\langle W_{\rm out}(\Omega)\rangle} \right]^2.
~\label{eq:TUR-work}
\end{equation}
(i) While the input driving energy is positive, i.e., $\langle W_I \rangle>0$, the free energy efficiency of the heat engine is~\cite{GingrichPRL}
\begin{equation}
\langle \eta \rangle = \frac{\langle W_{\rm out}\rangle}{\langle W_d \rangle+\langle W_I \rangle}= \frac{\langle W_{\rm out}\rangle}{T\langle \sigma \rangle+\langle W_{\rm out} \rangle}.
\end{equation}
According to the above definition of efficiency $\braket{\eta}$, the relation Eq.~\eqref{eq:TUR-work} implies

\begin{equation}
\frac{1}{\langle\eta\rangle} \ge 2T \frac{\langle W_{\rm out} \rangle}{\langle\langle W_{\rm out}^2 \rangle\rangle} \left[1 - \Omega\frac{d \langle W_{\rm out}(\Omega)\rangle/d\Omega}{\langle W_{\rm out}(\Omega)\rangle} \right]^2 +1.
~\label{eq:eta_W1}
\end{equation}
Furthermore, the output work contributions come from the dynamic part and the geometric one, i.e., $W_{\rm out}=W_{\rm out}|_{\rm dyn}+W_{\rm out}|_{\rm geo}$, and the geometric work $W_{\rm out}$ is proportional to the driven frequency $\Omega$. Then, the right-hand side of Eq.~\eqref{eq:eta_W1} can be simplified as
\begin{equation}
\begin{aligned}
\frac{1}{\langle\eta\rangle} \ge & 1 + 2T \frac{\langle W_{\rm out} \rangle}{\langle\langle W_{\rm out}^2 \rangle\rangle} \\
&\times \left[\frac{1}{1 + \langle W_{\rm out}(\Omega)\rangle|_{\rm geo} /\langle W_{\rm out}(\Omega)\rangle|_{\rm dyn}} \right]^2 \equiv \frac{1}{\eta_{\rm bound}},
\label{eq:TURp1}
\end{aligned}
\end{equation}
when $\langle W_I \rangle>0$.

(ii) While the driving energy is negative, i.e., $\langle W_I \rangle<0$, the free energy efficiency of the heat engine is~\cite{HinoPRR21,IzumidaPRE21},
\begin{equation}
\langle\eta\rangle = \frac{\langle W_{\rm out}\rangle}{\langle W_d \rangle}= \frac{\langle W_{\rm out}\rangle}{T\langle \sigma \rangle+\langle W_{\rm out} \rangle - \langle W_I \rangle}.
\end{equation}
Combined with the expression of the efficiency $\braket{\eta}$, the relation Eq.\eqref{eq:TUR-work} implies
\begin{equation}
\frac{1}{\langle\eta\rangle} \ge 2T \frac{\langle W_{\rm out} \rangle}{\langle\langle W_{\rm out}^2 \rangle\rangle} \left[1 - \Omega\frac{d\langle W_{\rm out}(\Omega)\rangle/d\Omega}{\langle W_{\rm out}(\Omega)\rangle} \right]^2 - \frac{\langle W_I \rangle}{\langle W_{\rm out} \rangle}+1.
~\label{eq:eta_W2}
\end{equation}
By further specifying two components in the output work, the Eq.~\eqref{eq:eta_W2} can be simplified as
\begin{equation}
\begin{aligned}
\frac{1}{\langle\eta\rangle} \ge &1 + 2T \frac{\langle W_{\rm out} \rangle}{\langle\langle W_{\rm out}^2 \rangle\rangle} \\
&\times \left[\frac{1}{1 + \langle W_{\rm out}(\Omega)\rangle |_{\rm geo}/\langle W_{\rm out}(\Omega)\rangle|_{\rm dyn}} \right]^2  \\
&- \frac{\langle W_I \rangle}{\langle W_{\rm out} \rangle|_{\rm dyn}+\langle W_{\rm out} \rangle|_{\rm geo}}\equiv \frac{1}{\eta_{\rm bound}}.
\label{eq:TURp2}
\end{aligned}
\end{equation}

Eqs.~\eqref{eq:TURp1} and \eqref{eq:TURp2} are our second main results.
In general, the output work of a steady-state heat engine vanishes at least linearly as its energy efficiency approaches unity~\cite{JiangPRE,JiangPRL,HolubecPRL18}. A finite power in this limit is, in principle, is possible only if the current fluctuations diverge~\cite{PietzonkaPRL} or if the output power is proportional to the cycling frequency of the engine~\cite{KoyukPRL19}. Although these general results have been demonstrated in previous studies, our work provides a general realizable optimization principle. By maximizing the geometric contribution $W_{\rm out}|_{\rm geo}$ using geometric methods and minimizing the dynamic contribution $W_{\rm out}|_{\rm dyn}$, we can push the bounds [Eqs.~\eqref{eq:TURp1} and \eqref{eq:TURp2}]  to a more efficient regime.

\begin{figure}[htb]
\begin{center}
\centering\includegraphics[width=8.5cm]{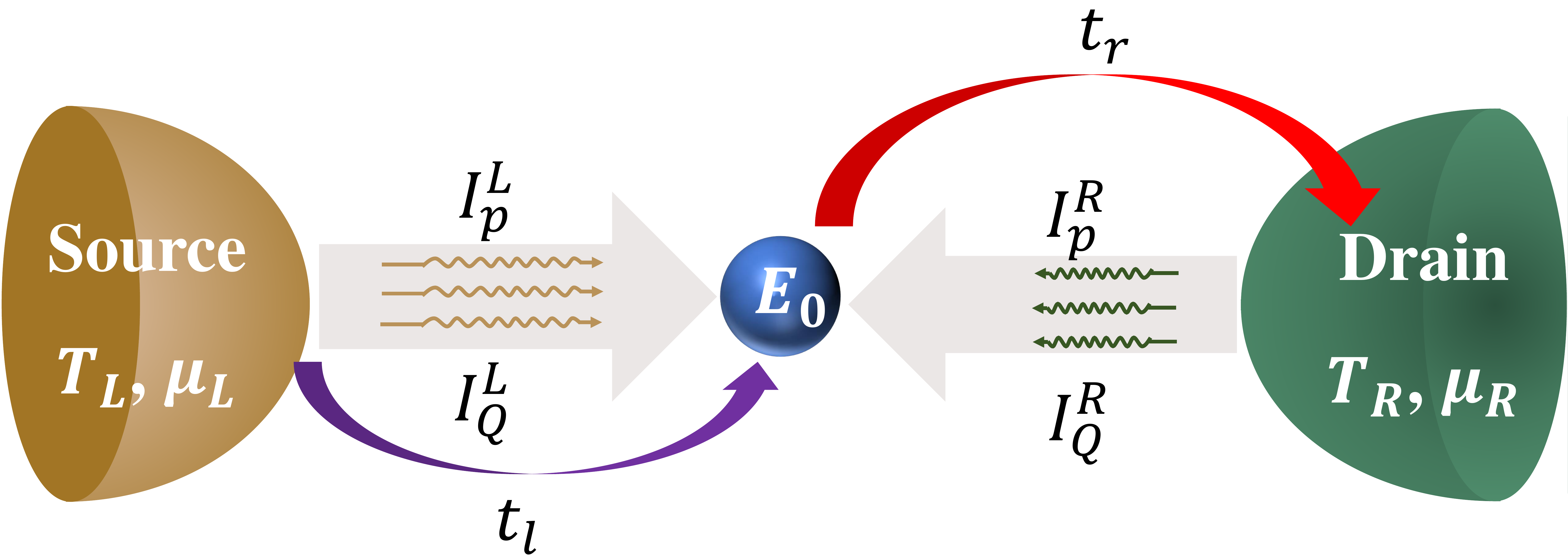}
\caption{ Schematic of the single-level QD system. An electron from the source (with chemical potential $\mu_L$ and temperature $T_L$) can flow across the QD (with energy $E_0$) and hop into the drain (with chemical potential $\mu_R$ and temperature $T_R$). The energy level in the QD, $E_0$, and the coupling $t_i$ ($i=l,r$) between the dot and the reservoir, can generally be taken as time-dependent driving parameters.}
\label{fig:2Tsystem}
\end{center}
\end{figure}

\section{Verifying the validity of the thermodynamic uncertainty relation in periodically driven thermoelectric
heat engine}\label{verify}

\subsection{Single-level quantum dot System}


We illustrate the formal results within the two-terminal system. In our construction (see Fig.~\ref{fig:2Tsystem}), a single quantum dot (QD) system is exchanging energy with two electronic reservoirs, $L$ and $R$, which can be set out of equilibrium with a finite voltage bias $\Delta\mu=\mu_R-\mu_L$ or/and  temperature difference ${\Delta}T=T_L-T_R$.
Our model is described by the Hamiltonian
\begin{equation}
\hat H = \hat H_{S} + \hat H_{B} + \hat H_{I},
~\label{total-Hami}
\end{equation}
where $ \hat H_{S} = E_0 c_d^{\dagger}c_d$ denotes the sigle-level QD, $\hat H_{B}=\sum_{v=L,R}\sum_k \epsilon_{kv} c_{kv}^{\dagger}c_{kv}$ represents the left and right electronic reservoirs (source and drain), and $\hat H_I=\sum_{v=L,R}\sum_k t_{kv}(c_{kv}^{\dagger}c_d + \rm {H.c.})$ is the system-reservoir interaction Hamiltonian.
The working substance consists of a single electronic level with the annihilation operator $c_d$ and time-dependent energy $E_0(t)$. The dot is alternating its coupling to two fermionic baths (leads) $v = L,R$, which may have different temperatures. $c_{kv}$ annihilates an electron with energy $\epsilon_{kv}$ in the $v$-lead that couples to the central level with $t_{kv}$ being the tunneling rate. The $v$-lead is characterized as the Fermi-Dirac distribution function
$f_v(\omega)=\{\exp[(\omega-\mu_v)/k_BT_v]+1\}^{-1}$, with an temperature $T_v$ and a chemical potential $\mu_v$. The reservoirs containing large number of states exert dissipative effects on the dynamics described by the spectral function $\Gamma_v(\epsilon)=2\pi\sum_kt_{kv}^2\delta(\epsilon-\epsilon_k)$.

Using the Redfield approximation for weak system-bath coupling~\cite{gardiner2004quantum,SegalPRB06,BijayJiang,JiangBijayPRB17}, the underlying dynamics can be modeled as
\begin{subequations}
\begin{align}
&\dot{p}_0^{\lambda}(t) = -k_{u} \, {p}^{\lambda}_{0}(t) +  k_{d}^{\lambda}  \,{p}^{\lambda}_1(t), \\
&\dot{p}_1^{\lambda}(t) =  k_{u}^{\lambda} \, {p}_0^{\lambda}(t)-k_{d} \,{p}_1^{\lambda}(t).
\end{align}
\end{subequations}
Here $\lambda$ is the counting parameter, which can be used to calculate the fluctuation properties of an arbitrary flow, heat, particle, etc., induced by quantum transitions.
These equations can be reexpressed in a matrix form as
\begin{equation}
\frac{d|p^{\lambda}(t)\rangle}{dt} = {\cal H}(\lambda) |p^{\lambda}(t)\rangle,
\end{equation}
where $|p^{\lambda}(t)\rangle = ({p}^{\lambda}_{0}(t),{p}^{\lambda}_{1}(t))$. And $p_n$ ($n=0,1$) denotes the probability of QD to occupy the state $|n\rangle$, satisfying $p_0(t)+p_1(t)=1$~\cite{SegalPRL08}. The activation and relaxation rates with the counting field read
\begin{subequations}
\begin{align}
k_{u}^{\lambda} &= k_{0\rightarrow 1}^L + k_{0\rightarrow 1}^R e^{i\lambda_p + iE_0\lambda_E},\\
k_{d}^{\lambda} &= k_{1\rightarrow 0}^L + k_{1\rightarrow 0}^R e^{-i\lambda_p - iE_0\lambda_E},
\end{align}
\end{subequations}
Here, $k_{0\rightarrow 1}^v=\Gamma_vf_v(E_0)$, and $k_{1\rightarrow 0}^v=\Gamma_v[1 - f_v(E_0)]$~\cite{Jiang2012,Jiangtransistors,MyPRBtransistor,MyPRBdemon}. $\lambda_E$ and $\lambda_p$ are the counting fields for energy and particles, respectively. Without loss of generality, here we count the flow between the system and the right reservoir. Here we define the positive current to be flowing from reservoirs into the middle system. 

Finally, the steady-state particle and energy currents flowing from the right reservoir into the system are expressed as    \cite{gchen2005book,jauho2008book}
\begin{subequations}
\begin{align}
&\braket{I_p^R}_s=\frac{\Gamma_L\Gamma_R[f_R(E_0) - f_L(E_0)]}{\Gamma_L+\Gamma_R},\\
&\braket{I_E^R}_s=\frac{E_0\Gamma_L\Gamma_R[f_R(E_0) - f_L(E_0)]}{\Gamma_L+\Gamma_R},
\end{align}
\end{subequations}
while $\braket{I_Q}$ is the net heat current carried by the electrons, $\braket{I_Q^R}=\braket{I_E^R}-\mu_R\braket{I_p^R}$. These steady state results are of the typical Landauer type in thermal transports~\cite{SegalPRB06}. The flows from the left and right reservoirs are not independent. Particle conservation implies that $\braket{I_p^L}_s+\braket{I_p^R}_s=0$, while energy conservation requires $\braket{I_E^L}_s+\braket{I_E^R}_s=0$~\cite{JiangCRP,MyPRBdiode}.



\begin{figure}[htb]
\begin{center}
\centering\includegraphics[width=8.5cm]{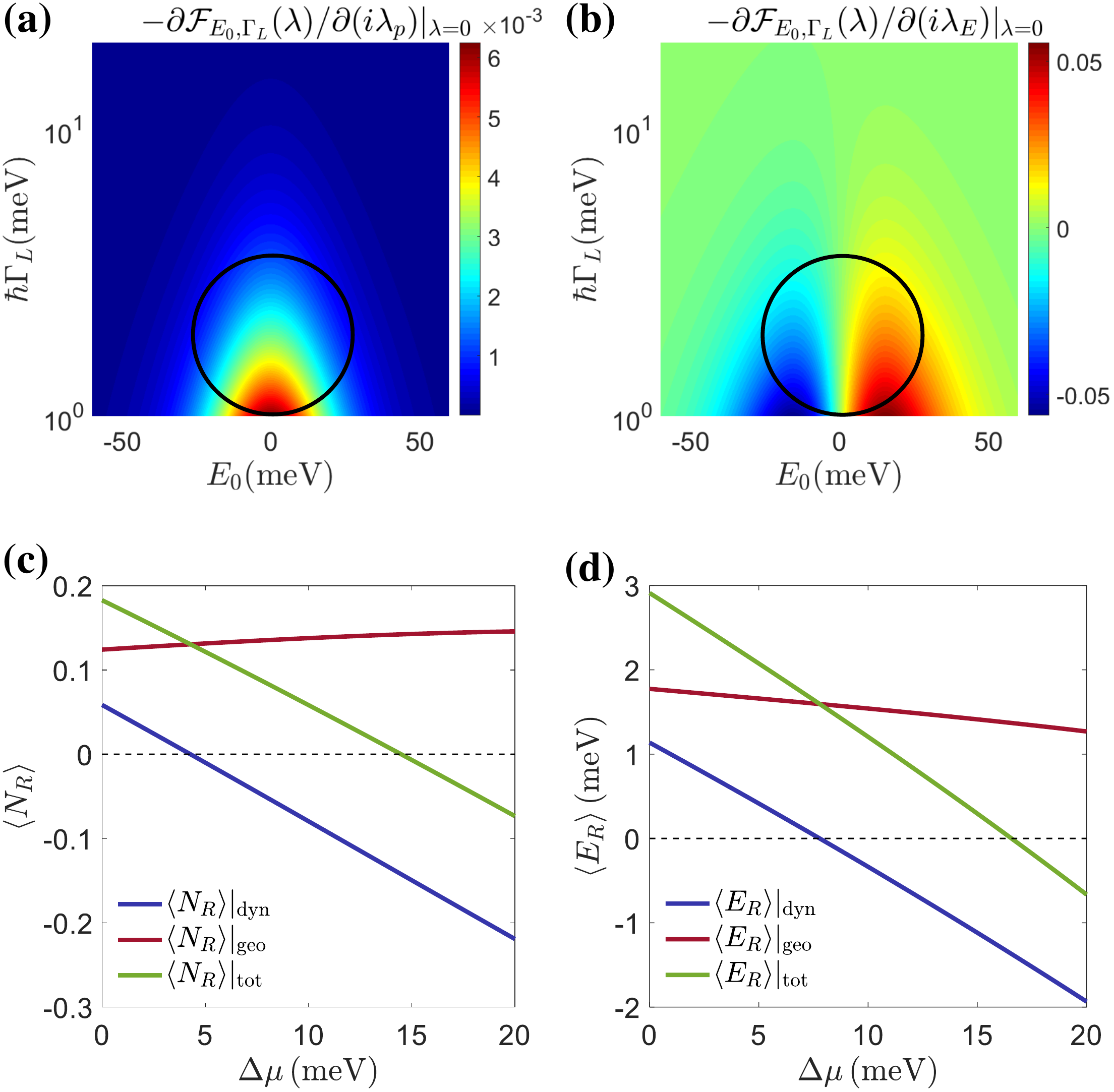}
\caption{The contour map of (a) Berry curvature for the average particle current: $-\partial{\mathcal F}_{E_0\Gamma_L}(\lambda)/\partial(i\lambda_p)|_{\lambda=0}$. (b) Berry curvature for the average energy current: $-\partial{\mathcal F}_{E_0\Gamma_L}(\lambda)/\partial(i\lambda_E)|_{\lambda=0}$, in the parameter space of the dot energy level $E_0$ and the left reservoir's spectral function $\Gamma_L$. (c) The particle current $N_R$. (d) The energy current $E_R$ as a function of $\Delta\mu$. Here, we define positive flows to be from the right to the left. The parameters are $\mu=0$ (mean value), $\hbar\Gamma_R=1 \, {\rm meV}$, $k_BT_L=10 \, {\rm meV}$ and $k_BT_R=1.5k_BT_L$. The energy modulations: $E_0= [15+15\cos(\Omega t)]\, {\rm meV}$, $\hbar\Gamma_L=[2 + \sin(\Omega t)]\, {\rm meV}$, $\Omega=2\pi/{\mathcal T}_p$ and ${\mathcal T}_p=10^{-12} \,{\rm s}$.} ~\label{fig:IEe-Deltamu}
\end{center}
\end{figure}

\subsection{Geometric Berry-phase-induced particle and energy currents}

For heat engine operation, the single QD system connected to the two reservoirs is subjected to cyclic parameter modulations. This could be realized by imposing a modulation on either of the following parameters: $\Gamma_v(t)$, $\mu_v(t)$, $T_v(t)$, ($v=L,R$) and $E_0(t)$~\cite{LudovicoPRB18,LvPRB21}. The particle and energy currents from the right ($R$) reservoir into the single QD system during the long time span $\tau$. The characteristic function is~\cite{RenPRL12}
\begin{equation}
{\mathcal Z}_{\tau} = \sum_{q=-\infty}^{+\infty} P_{\tau}(q)e^{iq\lambda}=1^{\dagger} \hat T [e^{\int_0^{\tau} {\mathcal H}(\lambda,t) dt}]{\mathbf p}(0).
\end{equation}
where $P_{\tau}(q)$ is the probability distribution of having current transferred from the $R$ reservoir into the single quantum dot system during time $\tau\rightarrow \infty$. Here $1^{\dagger}=[1,1]$, $\hat T$ denotes the time-ordering operator, and ${\mathbf p}(0)=[p_0(0),p_1(0)]^T$ are the initial occupation probabilities.

According to the large deviation principle and the adiabatic perturbation theory, the cumulant generating function are composed of two parts in the long time ($\tau$) limit~\cite{RenPRL10,RenPRL12},
\begin{equation}
\begin{aligned}
&{\mathcal Z}_{\tau} \approx e^{\tau {\mathcal G}}=e^{\tau({\mathcal G}_{\rm dyn}+{\mathcal G}_{\rm geo})},\\
&{\mathcal G}_{\rm dyn} ={\mathcal T}_p^{-1} \int_0^{{\mathcal T}_p}dt\chi(\lambda,t),\\
&{\mathcal G}_{\rm geo} =-{\mathcal T}_p^{-1} \int_0^{{\mathcal T}_p}dt \langle \varphi(\lambda)|\partial_t|\psi(\lambda)\rangle.
\end{aligned}
\end{equation}
Here $\chi$ denotes the eigenvalues of the evolution matrix ${\mathcal H}$ with the biggest real part. $|\psi({\lambda},t)\rangle$ and $\langle\varphi({\lambda},t)|$ are the corresponding normalized right and left eigenvector, respectively. Obviously, given a paramter path ${\mathcal G}_{\rm dyn}$ is independent of the driving frequency $\Omega$, while ${\mathcal G}_{\rm geom}$ has a factor ${\mathcal T}_p^{-1}$ and is thus proportional to $\Omega$. This argument solidifies the scaling properties of $\braket{I}|_{\rm dyn}$ and $\braket{I}|_{\rm geo}$ in our deriving TUR relations (Sec.~\ref{sec-bounds}).


The first contribution ${\mathcal G}_{\rm dyn}$ presents the temporal average and defines the dynamic particle and heat transfer. This is the only term which survives in the static limit. The second, geometric part ${\mathcal G}_{\rm geo}$ presents an additional contribution caused by the adiabatic cyclic evolution and it requires at least two parameter modulations. For the case of periodically driving pairs [$u_1(t)$, $u_2(t)$],  which could be chosen from quantum dot energy [$E_0(t)$] and tunneling rate [$\Gamma_L(t)$], we have
\begin{equation}
{\mathcal G}_{\rm geo} =-{\mathcal T}_p^{-1} \iint_{u_1u_2} du_1du_2 {\mathcal F}_{u_1u_2}(\lambda),
\end{equation}
\begin{equation}
{\mathcal F}_{u_1u_2} = \langle \partial_{u_1} \varphi|\partial_{u_2} \psi\rangle - \langle \partial_{u_2} \varphi|\partial_{u_1} \psi\rangle,
\end{equation}
where ${\mathcal F}_{u_1u_2}$ is  analogous with the gauge invariant Berry curvature~\cite{berry84,bohm03}. The particle current flowing from the right reservoir into the system emerges as
\begin{equation}
I_p^R(t) = \frac{\partial{(\mathcal G_{\rm dyn}+\mathcal G_{\rm geo}})}{\partial (i\lambda_p)}|_{\lambda=0},
\end{equation}
and the energy current is
\begin{equation}
I_E^R(t) = \frac{\partial{(\mathcal G_{\rm dyn}+\mathcal G_{\rm geo}})}{\partial (i\lambda_E)}|_{\lambda=0}.
\end{equation}
The electronic heat current extracted from the right reservoir is defined as $I_Q^R(t) = I_E^R(t) - \mu_R I_p^R(t)$. The particle current $I_p^L$ and energy current $I_E^L$ flowing from the left ($L$) reservoir into the central system can be achieved upon introducing the activation and relaxation rates with the counting field $k_{u}^{\lambda} = k_{0\rightarrow 1}^Le^{i\lambda_p + iE_0\lambda_E} + k_{0\rightarrow 1}^R$, $k_{d}^{\lambda} = k_{1\rightarrow 0}^Le^{-i\lambda_p - iE_0\lambda_E} + k_{1\rightarrow 0}^R$. The fluctuation of the current is $\braket{\braket{I^2}}={\partial^2\mathcal G}/{\partial (i \lambda)^2}|_{\lambda=0}$, where $\braket{\braket{I^2}}=\braket{I^2}-\braket{I}^2$ is the second cumulant.




We now turn to the first law of thermodynamics for time-dependent quantum thermal machine~\cite{jauho2008book}. The particle conversation is characterized as
\begin{equation}
\int_0^t d \tau [I_p^L(\tau)+I_p^R(\tau)]=\Delta n,
\label{eq:parcon}
\end{equation}
and the energy conversation is quantified by~\cite{qubitengine,wangpump,PRRes21,IzumidaPRE21,PRA21}
\begin{equation}
\int_0^t d\tau \left[\sum_{v=L,R}I_E^v(\tau)+A(\tau)\right]= \Delta U,
\label{eq:energycon}
\end{equation}
where $\Delta n$ and $\Delta U$ are respectively the stochastic occupation number and energy change of the middle system during $[0,t]$. The input power induced by temporal driving provides an extra term $A(t)$. Since terms in the left hand sides of Eq.~(\ref{eq:parcon}) and~(\ref{eq:energycon}) are growing with time and the right-hand sides are naturally bounded by the size of the middle system, we arrive at the approximate conservation laws in the long time (large period number) limit~\cite{ArracheaPRB20,JunjiePRL}
\begin{equation}
\begin{aligned}
\langle N_L \rangle + \langle N_R \rangle&=0,  \\
\braket{ E_L} + \langle E_R \rangle + \langle W_I \rangle &=0.
\label{eq:conversation}
\end{aligned}
\end{equation}
Here, $\langle N_v\rangle$ is the average accumulated input particle number into reservoir $v$, $\langle E_v\rangle$ is the input energy flowing from reservoir $v$ and $\langle W_I\rangle$ is the extra work done by driving. These stochastic quantities are accumulated during a given time interval.

We proceed by presenting simulation results at finite temperature and bias, focusing on the large-bias limit rather than the linear-response behavior. For convenience, We define the chemical potential difference $\Delta\mu=\mu_L-\mu_R$ and the mean chemical potential $\mu \equiv (\mu_L+\mu_R)/2$. We fix the mean chemical potential $\mu$ at zero, and study the effect of $\Delta \mu$. To operate the device as a thermoelectric engine, we assume $T_L<T_R$ and $\Delta \mu>0$. The produced electronic work after a period of driving cycle of the thermoelectric heat engine is~\cite{HinoPRR21,alonso21}
\begin{equation}
\langle W_{\rm out} \rangle= -(\mu_R-\mu_L) \langle N_R\rangle.
~\label{eq:power}
\end{equation}

In Fig.~\ref{fig:IEe-Deltamu}, we demonstrate the geometric thermoelectric pump effect. In Figs.~\ref{fig:IEe-Deltamu}(a) and \ref{fig:IEe-Deltamu}(b), we illustrate that by modulating parameters of the system (the system-reservoir coupling strength $\Gamma_L$, and the QD energy level $E_0$), the non-trivial Berry curvature leads to both non-vanishing geometric particle and energy flow, as shown in Figs.~\ref{fig:IEe-Deltamu}(c) and \ref{fig:IEe-Deltamu}(d). We note that the geometric flow currently can be experimentally observed~\cite{wang2021}, and our setup is within the reach of current experiment platforms~\cite{jaliel2019experimental,MailletPRL19,josefsson18}.

As a main result we demonstrate that the Berry-phase effect acts as a reconfigurable pump, providing additional particle and electronic heat currents across the QD systems with no static bias or even against the direction of biases. This shows the power of our framework. By tailoring the driving path in the parameter space, we can both design the functionality of thermoelectric engines, modulate the ratio of the geometric and the dynamic components of flows, and therefore optimize the engine's performance bounds proposed in Sec.~\ref{sec-bounds}. We elaborate this point in the following sections.


\begin{figure}[htb]
\begin{center}
\centering\includegraphics[width=8.5cm]{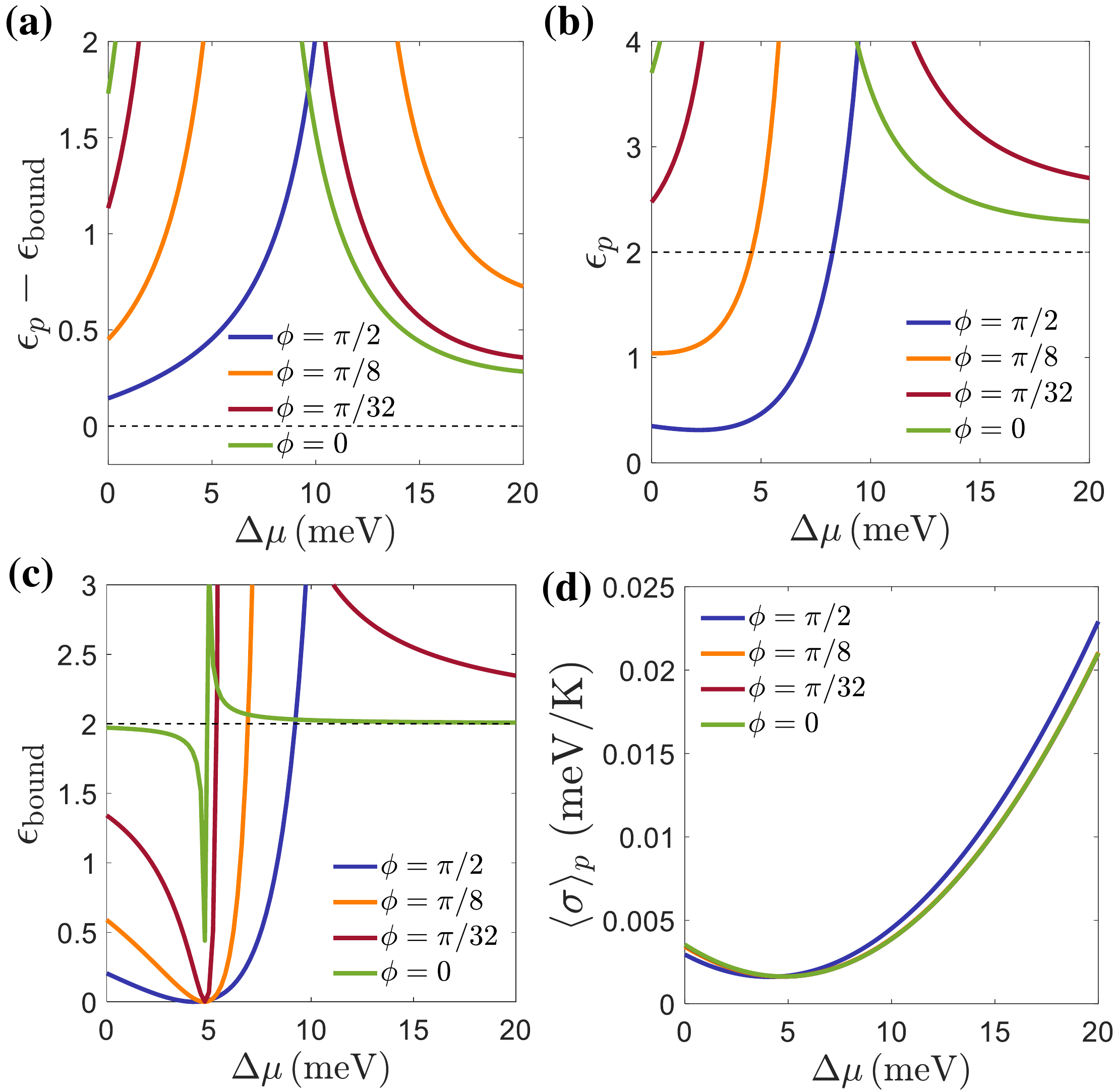}
\caption{(a) The distance of realized $\epsilon_p$ from the geometric TUR bound, i.e. $\epsilon_p-\epsilon_{\rm bound}$, (b) the original TUR bound $\epsilon_p$, (c) The bound $\epsilon_{\rm bound}$, (d) the average entropy production $\langle \sigma \rangle_p$ as a function of $\Delta\mu$ for different phase $\phi$. The parameters are $\mu=0$, $\hbar\Gamma_R=1 \, {\rm meV}$,  $k_BT_L=10 \, {\rm meV}$ and $k_BT_R=15 \, {\rm meV}$, $E_0= [15+15\sin(\Omega t+\phi)]\, {\rm meV}$, $\hbar\Gamma_L=[2 + \sin(\Omega t)]\, {\rm meV}$, $\Omega=2\pi/{\mathcal T}_p$ and ${\mathcal T}_p=10^{-12} \,{\rm s}$.}
~\label{fig:epsilon-p}
\end{center}
\end{figure}

\begin{figure}[htb]
\begin{center}
\centering\includegraphics[width=8.5cm]{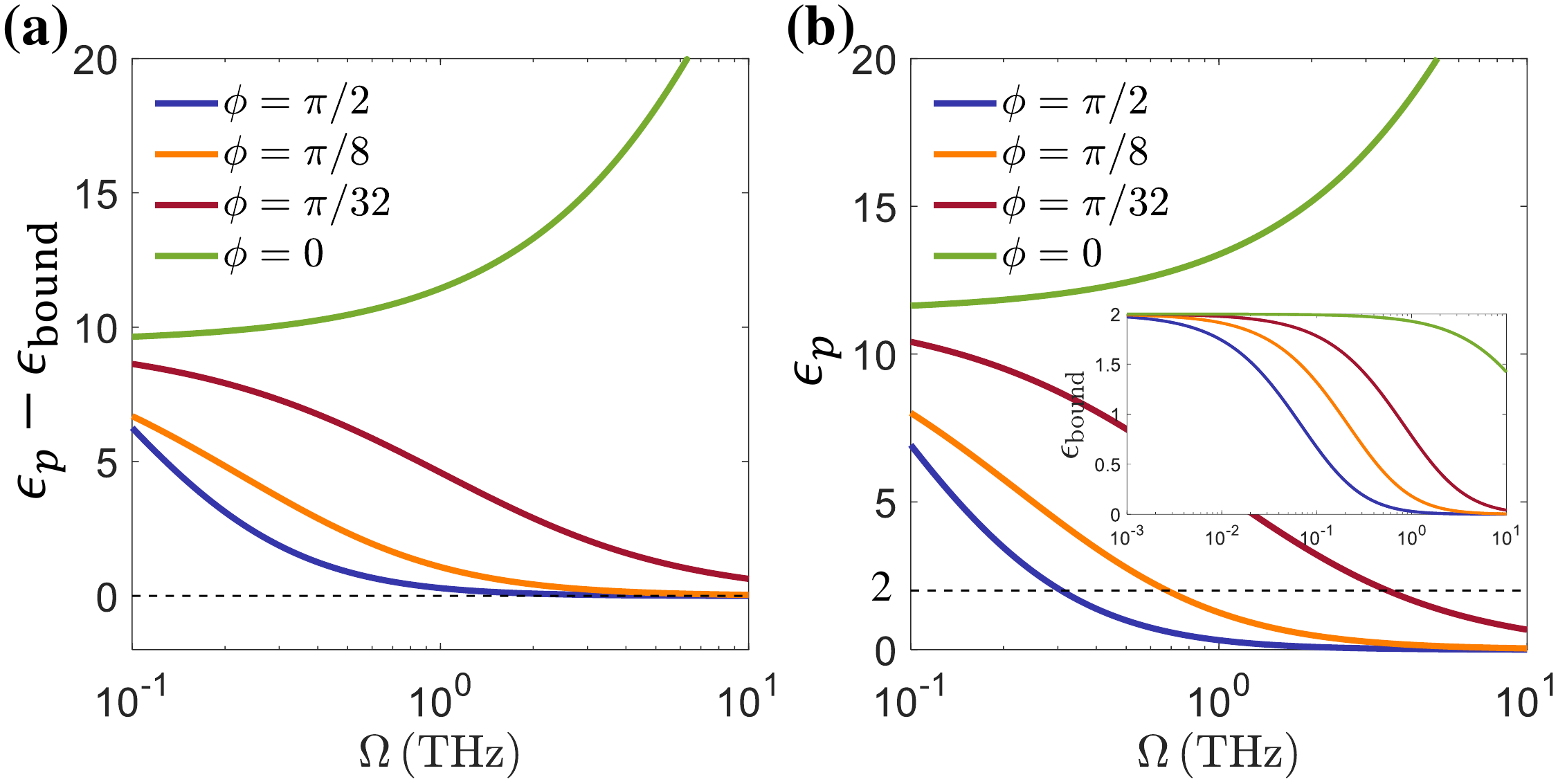}
\caption{(a) The distance of realized $\epsilon$ from the geometric TUR bound, i.e. $\epsilon_p-\epsilon_{\rm bound}$, and (b) the original TUR bound $\epsilon_p$, as a function of deriving frequency $\Omega$ for different phase $\phi$. Inset: the geometric TUR bound $\epsilon_{\rm bound}$ as a function of deriving frequency $\Omega$. The parameters are $\Delta\mu=3\, {\rm meV}$, $\mu=0$, $\hbar\Gamma_R=1 \, {\rm meV}$,  $k_BT_L=10 \, {\rm meV}$ and $k_BT_R=15 \, {\rm meV}$, $E_0= [15+15\sin(\Omega t+\phi)]\, {\rm meV}$, $\hbar\Gamma_L=[2 + \sin(\Omega t)]\, {\rm meV}$, and $\Omega=2\pi/{\mathcal T}_p$.}\label{fig:TUR-Omega}
\end{center}
\end{figure}

\subsection{Verifying the validity of the TUR relationship}

In Figs.~\ref{fig:epsilon-p} and \ref{fig:TUR-Omega}, we verify the bounds on fluctuations and entropy production in periodically driven systems. We concentrate on the effect geometric properties, which can be characterized by the relative phases $\phi$. Different $\phi$ represents different driving protocols. Specifically, parameters are driven as $[u_1(\Omega t),u_2(\Omega t+\phi)]$, with $u_1(\Omega t)$ and $u_2(\Omega t)$ being in phase. $\phi=\pi/2$ is the situation where the geometric contribution is optimized; in contrast, if the phase $\phi=0$, the geometric contribution vanishes and there is {\it only} the dynamic one. This is obvious since the encircled area in the parameter space vanishes if $\phi=0$.

In this work, we focus our discussions on the working regime of periodically driven thermoelectric heat engine, i.e., $\braket{W_{\rm out}}>0$. To assess our derived general bound, we denote the relative fluctuation as
\begin{eqnarray}~\label{epsp}
\epsilon_p \equiv \braket{\braket{W_{\rm out}^2}}\braket{\sigma}/\braket{W_{\rm out}}^2.
\end{eqnarray}
Here, $W_{\rm out}$ is the useful output work of the stochastic thermoelectric work. From Figs.~\ref{fig:epsilon-p}(a) and \ref{fig:TUR-Omega}(a), we find that, regardless of phase $\phi$, voltage bias $\Delta\mu$, and the driving frequency $\Omega$, the proposed geometric TUR bound $\epsilon_p-\epsilon_{\rm bound}$, derived from Eq.~\eqref{eq:TURp}, is always greater than zero. The bounds on fluctuations and entropy production are always satisfied. Moreover, as shown in Fig.~\ref{fig:epsilon-p}(b), if the geometric current vanishes ($\phi=0$), the steady-state TUR Eq.~\eqref{eq:TUR} holds.
This is consistent with steady-state transport for classical Markov processes~\cite{TUR15,GingrichPRL}.
Likewise, the inset figure of Fig.~\ref{fig:TUR-Omega}(b) illustrates that in the adiabatic limit $\Omega \rightarrow 0$, $\braket{I}|_{\rm geo} / \braket{I}|_{\rm dyn}\rightarrow 0$, reducing our TURs to the steady-state result.
However, as the phase $\phi$ becomes finite, e.g., $\phi=\pi/8$ and $\phi=\pi/2$, the geometric contribution dominates the heat transport and fluctuations.
Accordingly, the original TUR bound breaks down, since $\epsilon_p$ is lesser than 2 for the small voltage bias.
In sharp contrast, our geometric TUR is still robust.
Furthermore, in the regime of high driving frequency, the geometric-phase effect is pronounced, and it exactly the reason for breaking steady-state TUR [Eq.~\ref{eq:TUR}] and enhancing the precision of engines ($\epsilon_p<2$).


Then, we consider harvesting the heat from the (hot) reservoir and the energy for regulating QD system to generate electricity.
The entropy production of the whole system is $\langle \sigma \rangle_p=-\sum_{v=L,R} \langle Q_v\rangle/T_v$~\cite{e18110419}.
Considering the energy and particle conversations [Eq.~\eqref{eq:conversation}], the entropy production is given by a specific form of Eq.~(\ref{eq:entropy production})
\begin{equation}
\begin{aligned}
T_L\langle \sigma \rangle_p = -\langle W_{\rm out}\rangle+\left(1- {T_L/T_R}\right)\langle Q_R\rangle + \langle W_I\rangle.
\end{aligned}
\end{equation}

The thermal machine can be operated as a heat engine, the electric power $\langle W_{\rm out}\rangle>0$. If the input energy is negative, i.e., $\langle W_I\rangle<0$, the free energy efficiency of the heat engine is specified as~\cite{JiangJAP,ArracheaPRB16,ArracheaPRB16added,alonso21}
\begin{equation}
\langle \eta \rangle_p= \frac{\langle W_{\rm out}\rangle}{(1 - {T_L/T_R})\langle Q_R\rangle},
\end{equation}
which is consistent with the energy efficiency of steady-state thermoelectric transport~\cite{MyJAP,Jiang2017}. While the input energy is positive, i.e., $W_I>0$, the free energy efficiency of the heat engine is obtained as~\cite{HinoPRR21}
\begin{equation}
\langle\eta \rangle_p= \frac{\langle W_{\rm out}\rangle}{(1 - {T_L/T_R})\langle Q_R\rangle+\langle W_I\rangle}.
\end{equation}
According to the thermodynamic second law, the thermoelectric engine efficiency is upper bounded, i.e., $\langle\eta\rangle_p \le 1$~\cite{jauho2008book}.
 However, we note that the information on the geometric effect can refine this bound. 

\begin{figure}[htb]
\begin{center}
\centering\includegraphics[width=8.5cm]{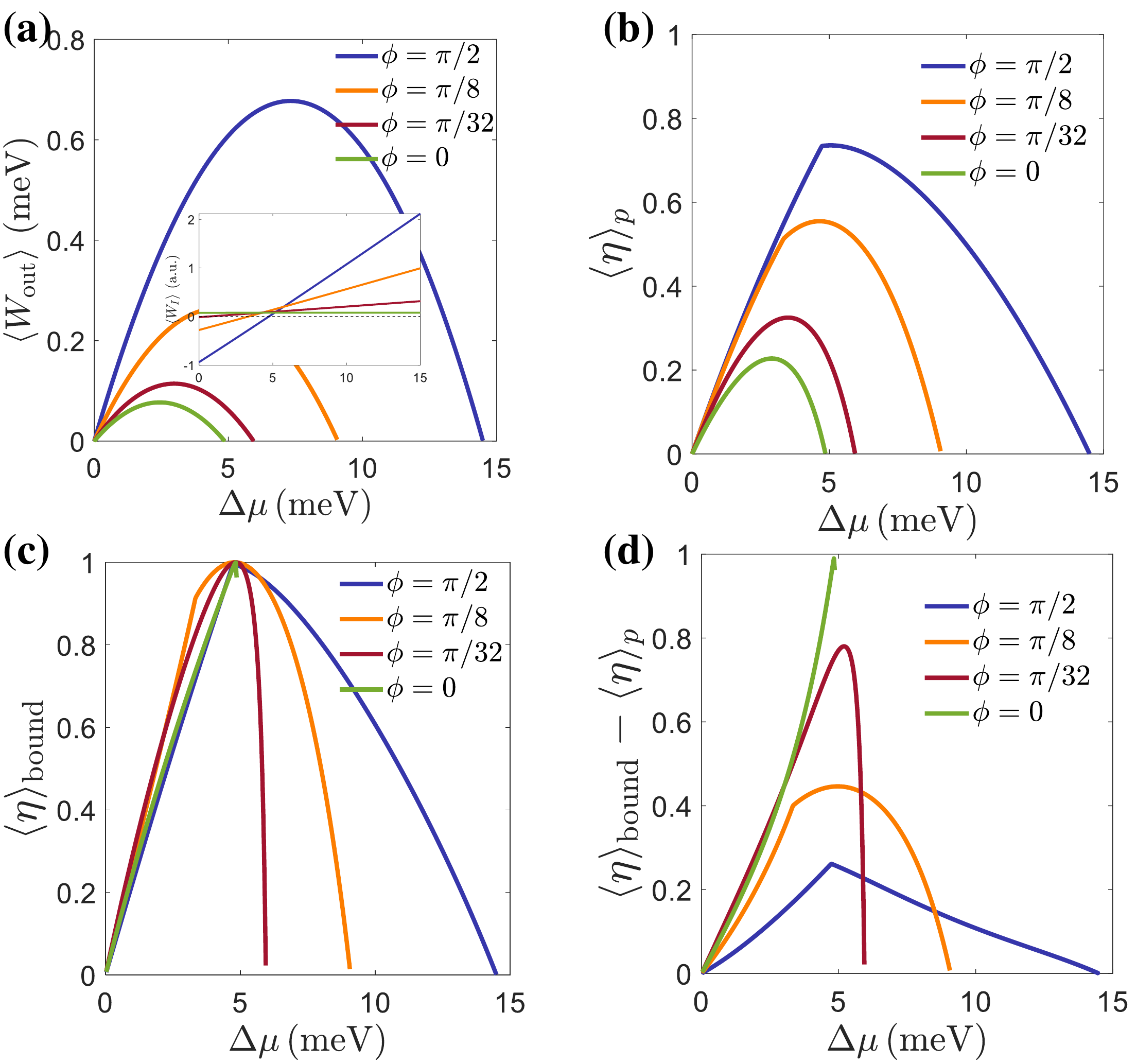}
\caption{(a) The output electric work $\braket{W_{\rm out}}$, (b) The energy efficiency $\braket{\eta}_p$, (c) $\langle \eta \rangle_{\rm bound}$, and (d) $\langle \eta \rangle_{\rm bound}-\langle \eta \rangle_p$ as a function of $\Delta\mu$ for different $\phi$. The parameters are $\mu=0$, $\hbar\Gamma_R=1 \, {\rm meV}$,  $k_BT_L=10 \, {\rm meV}$, $k_BT_R=1.5k_BT_L$, $E_0= [15+15\sin(\Omega t+\phi)]\, {\rm meV}$, $\hbar\Gamma_L=[2 + \sin(\Omega t)]\, {\rm meV}$, $\Omega=2\pi/{\mathcal T}_p$ and ${\mathcal T}_p=10^{-12} \,{\rm s}$.}
\label{fig:Peta_mu}
\end{center}
\end{figure}



\begin{figure}[htb]
\begin{center}
\centering\includegraphics[width=8.5cm]{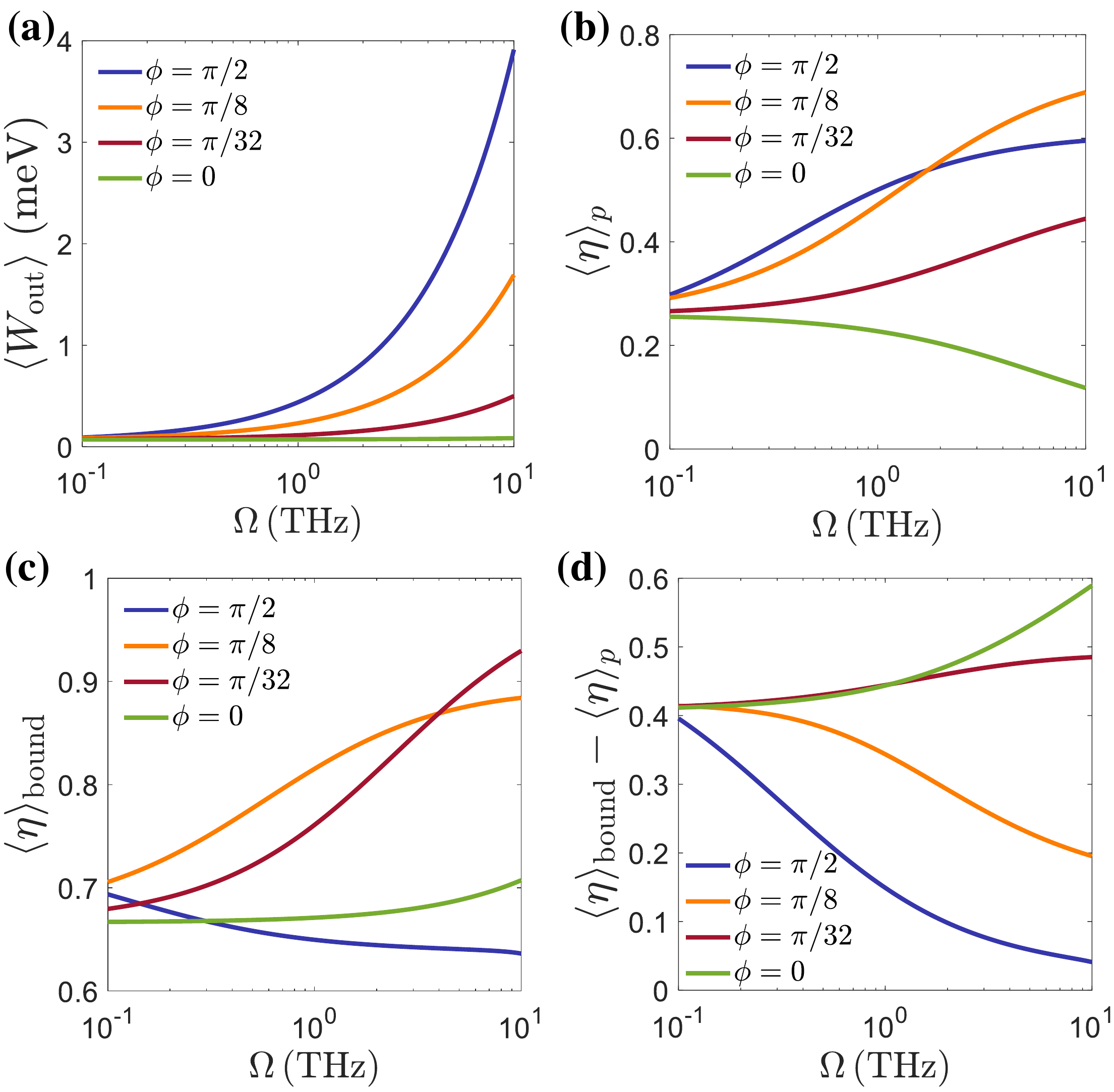}
\caption{(a) The output electric work $\braket{W_{\rm out}}$, (b) the energy efficiency $\braket{\eta}_p$, (c) $\langle \eta \rangle_{\rm bound}$, and (d) $\langle \eta \rangle_{\rm bound}-\langle \eta \rangle_p$ as a function of deriving frequency $\Omega$ for different phase $\phi$. The parameters are $\Delta\mu=3\, {\rm meV}$, $\mu=0$, $\hbar\Gamma_R=1 \, {\rm meV}$,  $k_BT_L=10 \, {\rm meV}$ and $k_BT_R=15 \, {\rm meV}$, $E_0= [15+15\sin(\Omega t+\phi)]\, {\rm meV}$, $\hbar\Gamma_L=[2 + \sin(\Omega t)]\, {\rm meV}$, and $\Omega=2\pi/{\mathcal T}_p$.}
\label{fig:Peta-omega}
\end{center}
\end{figure}

We now illustrate the geometric bound on efficiency in Eqs.~\eqref{eq:eta_W1} and \eqref{eq:eta_W2}. As shown in Fig. \ref{fig:Peta_mu}(a) and \ref{fig:Peta-omega}(a), we respectively illustrate the effect of voltage bias $\Delta\mu$ and driving frequency $\Omega$ on the electric work $\braket{W_{\rm out}}$ per driving period. Similarly, we also show $\braket{\eta}_p$ in Figs. \ref{fig:Peta_mu}(b) and \ref{fig:Peta-omega}(b). Obviously, the geometric phase yields significant improvement of the maximum efficiency and output work. From the efficiency change curve of Fig. \ref{fig:Peta_mu}(b), we find there is a turning point when the efficiency nearly reaches the summit. The reason for this phenomenon is that the input energy of the driving system changes from output to input, that is, the symbol of $\braket{W_I}$ is changed, which is illustrated in the inset of Fig. \ref{fig:Peta_mu}(a).

In Figs.~\ref{fig:Peta_mu}(c-d) and Figs.~\ref{fig:Peta-omega}(c-d),  we plot  efficiency bound $\eta_{\rm bound}$ and their difference $\eta_{\rm bound}-\langle \eta \rangle_p$ as a function of voltage bias $\Delta\mu$ and deriving frequency $\Omega$ for a thermoelectric engine, respectively. In parameter regime of the heat engine, the energy efficiency $\langle \eta \rangle_p$ never break through the boundary $\eta_{\rm bound}$, these simulations exemplify the validity of the geometric bounds. Interestingly, as shown in Figs.~\ref{fig:Peta_mu}(d) and ~\ref{fig:Peta-omega}(d), the efficiency comes even closer to its bound, i.e., $\eta_{\rm bound}-\braket{\eta}_p$ approaching zero, if we optimize the geometric effect  via  increasing either the driving frequency or $\phi$.

Finally, we discuss the thermodynamic bounds  of $\epsilon_p$ and the efficiency $\braket{\eta}_p$ in driven systems, compared to the steady-state ones.
For $\epsilon_p$ in Eq.~(\ref{epsp}), it is always greater than the lower bound $\epsilon_{\rm bound} \equiv 2/[1+{{\langle}W_{\textrm{out}}{\rangle}|_{\textrm{geo}}}/{{\langle}W_{\textrm{out}}{\rangle}|_{\textrm{dyn}}}]^2$.
Hence, once the geometric-phase-induced output work emerges, i.e. ${\langle}W_{\textrm{out}}{\rangle}|_{\textrm{geo}}{\neq}0$,
the lower limit of $\epsilon_p$ is generally unequal to the static counterpart $2$, which is also exhibited in Fig.~\ref{fig:epsilon-p}(c).
Interestingly, as the geometric component exceeds the dynamic one, the lower bound may even approach $0$, e.g., $\phi=\pi/8$.
Such picture can be alternatively explained that  though the output work is strongly enhanced by the geometric effect [Fig.~\ref{fig:Peta-omega}(a)],
the entropy production is not significantly increased [Fig.~\ref{fig:epsilon-p}(d)]. Similarly, $\braket{\eta}_{\rm bound}-\braket{\eta}_p \ge 0$ is always satisfied,
with the upper bound $\braket{\eta}_{\rm bound}$ given in Eq.~(\ref{eq:TURp1}) and Eq.~(\ref{eq:TURp2}) for
${\langle}W_{I}{\rangle}>0$ and ${\langle}W_{I}{\rangle}<0$, respectively.
This is also illustrated in Figs.~\ref{fig:Peta_mu}(d) and \ref{fig:Peta-omega}(d).
In the positive input energy generation regime, as the geometric-phase-induced work dominates the output work, i.e.,
${\langle}W_{\textrm{out}}{\rangle}|_{\textrm{geo}}{\gg}{\langle}W_{\textrm{out}}{\rangle}|_{\textrm{dyn}}$,
the entropy production ${\langle}\sigma{\rangle}$ is comparatively negligible to ${\langle}W_{\textrm{out}}{\rangle}|_{\textrm{geo}}/T$,
which results in the upper bound approaching the unity.
While for the case of negative input energy,
the efficiency bound can also reach the unity by further considering ${\langle}W_I{\rangle}{\ll}{\langle}W_{\textrm{out}}{\rangle}|_{\textrm{geo}}$.
 Hence, this shows the significance of geometric-phase in bounding the efficiency of heat engines by  observing the fluctuation of output work. Moreover, such results are in agreement with comparable counterparts in the stochastic clock~\cite{BaratoPRX}.
Therefore, we conclude that geometric part contribution is incredible to dramatically modify the TUR.

\section{conclusions}\label{conclusion}

In summary, for periodically driven systems, we have proposed a class of inequalities, termed as geometric TUR, that relate the entropy production, with the mean of current and its variance by bringing to light the Berry-phase-like effect. This leads to a general trade-off relation between the output work, effective efficiency, entropy production, and an external control protocol. The corresponding bounds indicate that the geometric phase plays a key role in constraining the relative fluctuation of currents. Moreover, such bounds provide insight into the understanding of the precision of thermoelectric heat engine.
We note that our theory is able to be applied to systems arbitrarily far from equilibrium, and does not assume any specific symmetry of the system. To demonstrate the practical applicability of our results, we work out the example of a two-terminal single level QD system, which lies within the family of thermoelectric heat engine. Our work paves the way for TUR from the geometric origin and optimizing more complex periodically driven thermoelectric heat engines.

\section{Acknowledgements}
J.L., Z.W., J.P., and J.R. acknowledge the support by the National Natural Science Foundation of China (Grant Nos. 11935010 and 11775159), the Natural Science Foundation of
Shanghai (Grant Nos. 18ZR1442800 and 18JC1410900). C.W. aknowledges support from the National Natural Science Foundation of China (Grant No. 11704093) and the Opening Project of Shanghai Key Laboratory of Special Artificial Microstructure Materials and Technology. J.-H.J acknowledges support from  the Natural Science Foundation of China (NSFC) (Grant Nos. 12074281, 12047541, and 12074279), the Major Program of Natural Science Research of Jiangsu Higher Education Institutions (Grant No. 18KJA140003), the Jiangsu specially appointed professor funding, and the Academic Program Development of Jiangsu Higher Education (PAPD).

\bibliography{Ref-pumping}

\end{document}